\documentclass[final,1p,times]{elsarticle} 

\usepackage{graphicx}
\usepackage{amssymb} 
\usepackage{amsthm} 
\usepackage{lineno}
\usepackage{wrapfig}
\usepackage{caption}
\usepackage{subcaption}

\journal{Nuclear Physics A} 

\begin{document}
\begin{frontmatter} 

\title{STAR Results from the RHIC Beam Energy Scan-I}

\author{Lokesh Kumar (for the STAR\fnref{col1} Collaboration)}
\fntext[col1] {A list of members of the STAR Collaboration and acknowledgements can be found at the end of this issue.}
\address{Department of Physics, Kent State University, Kent, OH 44242}


\begin{abstract} 
The Beam Energy Scan (BES) program is being pursued at RHIC to study the
QCD phase diagram, 
and
search for the possible QCD phase boundary and possible QCD
critical point. The data for Phase-I of the BES program 
have been collected for  
Au+Au collisions 
at center-of-mass energies
($\sqrt{s_{NN}}$) of  7.7, 11.5, 19.6, 27, and 39 GeV. 
These collision energies allowed the STAR experiment to 
cover a wide range of baryon chemical potential $\mu_{B}$ (100--400 MeV)
in the QCD phase diagram. We report on several
interesting results from the BES Phase-I 
covering the high net-baryon density region. These results 
shed light on particle production mechanism and freeze-out conditions,
first-order phase transition and ``turn-off'' of QGP signatures, and
existence of a critical point in the phase diagram. Finally, 
we give an outlook for the future BES Phase-II program and a possible
fixed target program at STAR.
\end{abstract} 

\end{frontmatter} 


\section{Introduction}
The results from top RHIC energies suggest the existence of the Quark
Gluon Plasma (QGP)~\cite{qgp}. The main task at hand now is to
study the properties of the QGP and establish the QCD phase
diagram. Lattice QCD calculations predict the transition between QGP and
the hadronic gas as crossover at $\mu_{B}=$ 0, 
while at large $\mu_{B}$ they predict a first order phase
transition~\cite{lattice1}. 
A point where the first order phase transition ends is
called the critical point.
Experimentally, 
the QCD phase diagram can be studied  by colliding heavy ions at varying beam energies that
can provide a $T$-$\mu_{B}$ region for each energy. Then, one can
look at the various signatures for the phase boundary and
critical point. 
We present results on (a) freeze-out parameters to get 
insight 
into the 
QCD phase diagram, (b) 
first and second coefficients ($v_1$, $v_2$) of the Fourier expansion of the angular distributions,
dynamical charge correlations, and nuclear
modification factor $R_{\rm{CP}}$ for the search of  first-order phase
transition and ``turn-off'' of QGP signatures, and (c) fluctuation
measurements for the search of QCD critical point. 

The first STAR proposal for the BES program~\cite{ref_bes} was made in the year
2008.~This was followed by a successful data taking and physics analysis of a test run below injection energies at $\sqrt{s_{NN}}$ = 9.2 GeV~\cite{9gev}.
The first phase of the BES program was started in
the year 2010 with data taking at three low energies of 7.7, 11.5, and
39 GeV. In the year 2011, two more energy points were added at
$\sqrt{s_{NN}}=$ 19.6 and 27 GeV. 

The results presented here are based on the data collected by
the STAR detector. STAR covers a large acceptance of 2$\pi$ in azimuth ($\phi$)
and $-$1 to 1 in pseudorapidity ($\eta$). One of the important
advantages which STAR possesses for the BES program is its almost uniform
acceptance for different identified particles and collision energies at midrapidity.
The main tracking device at STAR is the Time Projection
Chamber (TPC), which provides momentum as well as particle
identification (PID).
For higher transverse momentum ($p_{T}$) region,
the Time Of Flight (TOF) detector is quite effective in 
distinguishing different particles. Particles are identified using the
ionization energy loss in TPC and time-of-flight information from TOF~\cite{tpc_tof}.
The centrality selection is done
in STAR using the uncorrected charge particle multiplicity measured in
the TPC within $|\eta|<$ 0.5~\cite{9gev}. 

\section{Accessing QCD Phase Diagram}
\begin{figure}[htbp]
\begin{center}
\vspace{-0.3cm}
\includegraphics[width=0.45\textwidth]{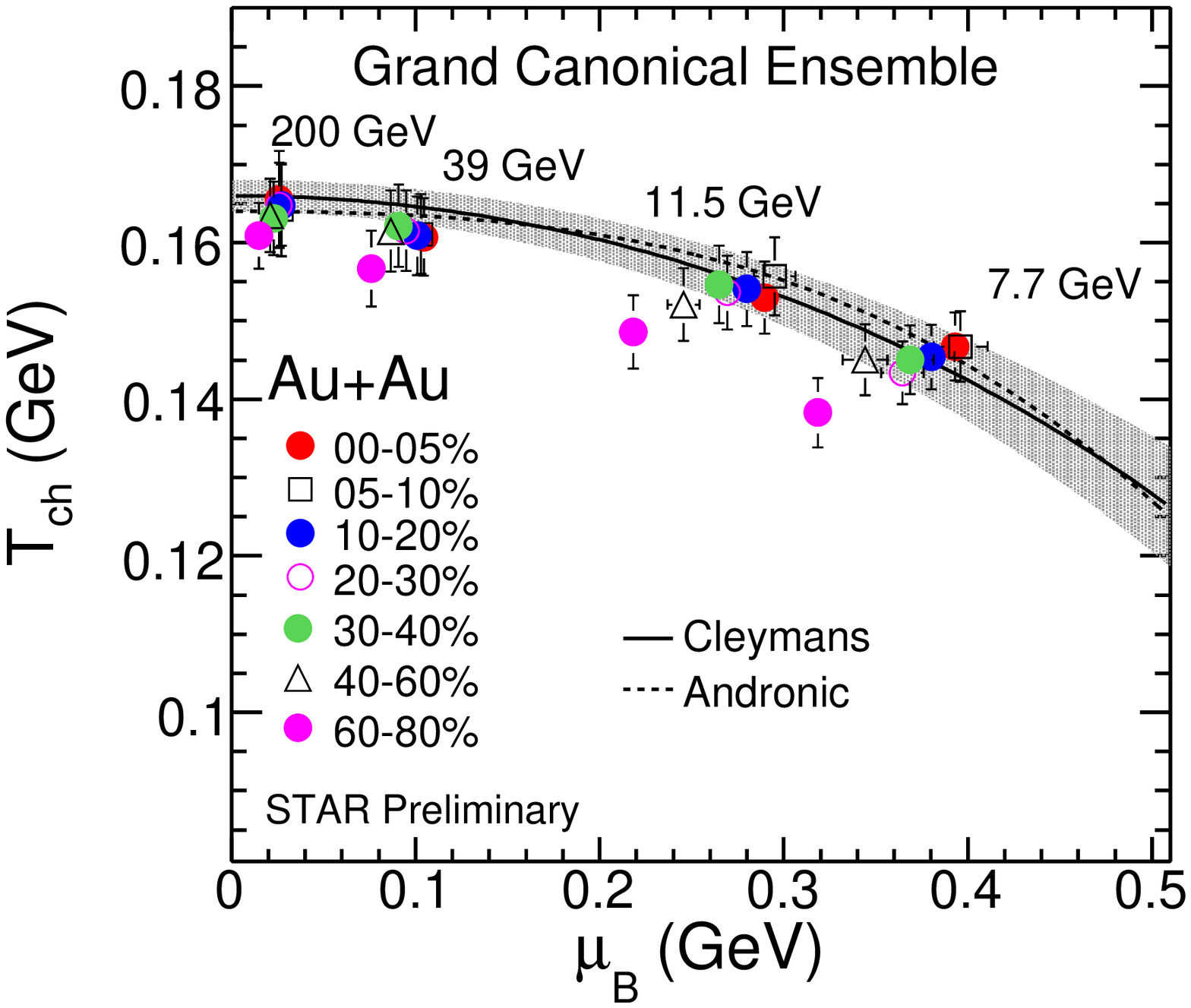}
\includegraphics[width=0.45\textwidth]{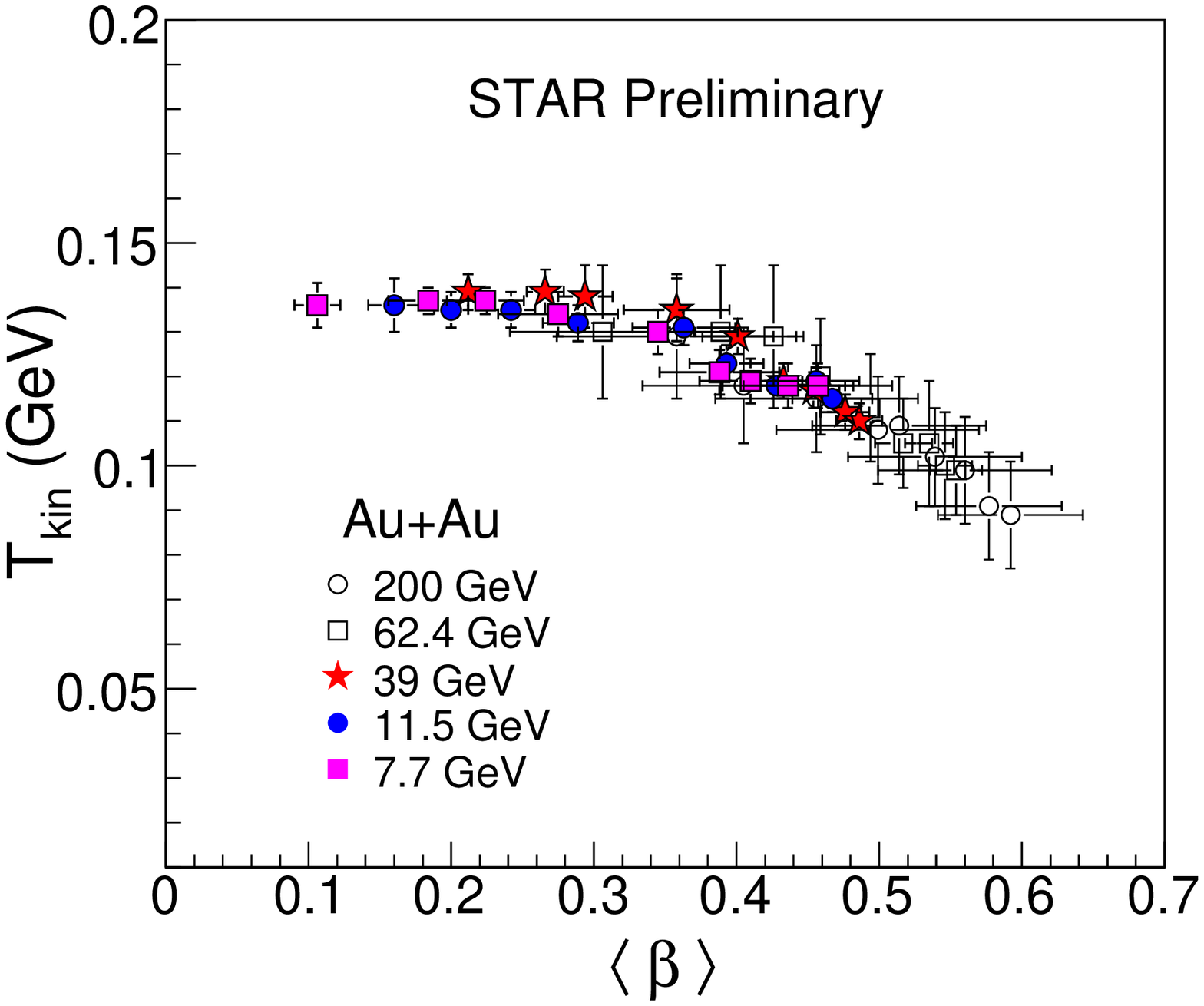}
\end{center}
\vspace{-0.75cm}
\caption{(Color online) 
Left panel: Variation of $T_{\rm{ch}}$ with $\mu_{B}$ for
different energies and centralities. 
The curves represent the theoretical calculations~\cite{fo_curves,cleymans}.
Right panel: Variation of
$T_{\rm{kin}}$ with $\langle \beta \rangle$ for different energies
and centralities. 
Errors in both panels represent the quadrature
sum of systematic and statistical errors.
}
\label{chem_kin}
\end{figure}

The QCD phase diagram is the variation of temperature $T$ and baryon
chemical potential $\mu_B$.
These quantities can be 
extracted from the measured hadron yields.
Transverse momentum spectra for the BES Phase-I energies are
obtained for $\pi$, $K$, $p$, $\Lambda$, $\Xi$, $K^{0}_{S}$, and
$\phi$~\cite{lok}.
The particle ratios are used to obtain the chemical freeze-out (a state
when the yields of particles get fixed)
conditions using the statistical thermal model
(THERMUS)~\cite{thermus}. The two main extracted parameters are chemical
freeze-out temperature $T_{\rm{ch}}$ and $\mu_{B}$. 
Figure~\ref{chem_kin} (left panel) shows the variation of the extracted
chemical freeze-out parameters using the Grand-Canonical Ensemble (GCE) approach of
THERMUS for different energies and centralities~\cite{sabita}.
We observe that at
lower energies, $T_{\rm{ch}}$ shows a variation with $\mu_{B}$ as a
function of centrality. Both $T_{\rm{ch}}$ and $\mu_{B}$ values seem to be
decreasing from central to peripheral collisions. It may be noted that
use of Strangeness Canonical Ensemble (SCE) shows an opposite trend,
i.e. the $T_{\rm{ch}}$ values seem to increase from central to
peripheral collisions~\cite{sabita}. However, it is found that the $\chi^{2}/NDF$ in
case of SCE is higher for peripheral collisions~\cite{sabita}. 
More investigations are ongoing. 
In GCE the energy and quantum numbers or particle numbers are
conserved on average through the temperature and chemical
potentials. GCE is widely used in heavy-ion collisions. In the SCE,
the strangeness ($S$) in the system is fixed exactly
by its initial value of $S,$ while the baryon and charge contents are treated grand-canonically.
Although the $T_{\rm{ch}}$ shows an opposite trend between GCE and SCE, both results suggest the variation of chemical
freeze-out parameters with centrality at lower energies. 

The particle spectra can be used to obtain the kinetic freeze-out
(a state
when the spectral shapes of particles get fixed)
conditions using the Blast Wave (BW) model~\cite{bw}. The BW model
is used to simultaneously fit the $\pi$, $K$, $p$ spectra and the two
relevant extracted parameters are kinetic freeze-out temperature $T_{\rm{kin}}$
and average flow velocity $\langle \beta
\rangle$. Figure~\ref{chem_kin} (right panel) shows the variation of
kinetic freeze-out parameters for different energies and
centralities~\cite{sabita}. 
We observe that at a given collision energy, there is an anti-correlation
between $T_{\rm{kin}}$ and $\langle \beta \rangle$. For a given
collision centrality, the freeze-out temperature at high energy is lower and the
average collectivity velocity $\langle \beta \rangle$ is larger due to expansion. 

\section{Search for First Order Phase Transition \& Turn-off of QGP Signatures}
\subsection{Directed Flow}
The directed flow $v_1$ is calculated as $\langle \cos(\phi-\Psi_1)
\rangle$, where $\phi$ and $\Psi_1$ are 
the azimuthal angle of the produced particles and orientation of the
first-order event plane, respectively. 
The directed flow measurements near midrapidity for protons are
considered as sensitive to the equation of state (EOS) and 
hence can be considered a phase transition signal~\cite{v1_ref}.
\begin{figure}[htbp]
\begin{minipage}{0.48\linewidth}
\centering
\vspace{-0.38cm}
\includegraphics[width=0.9\textwidth]{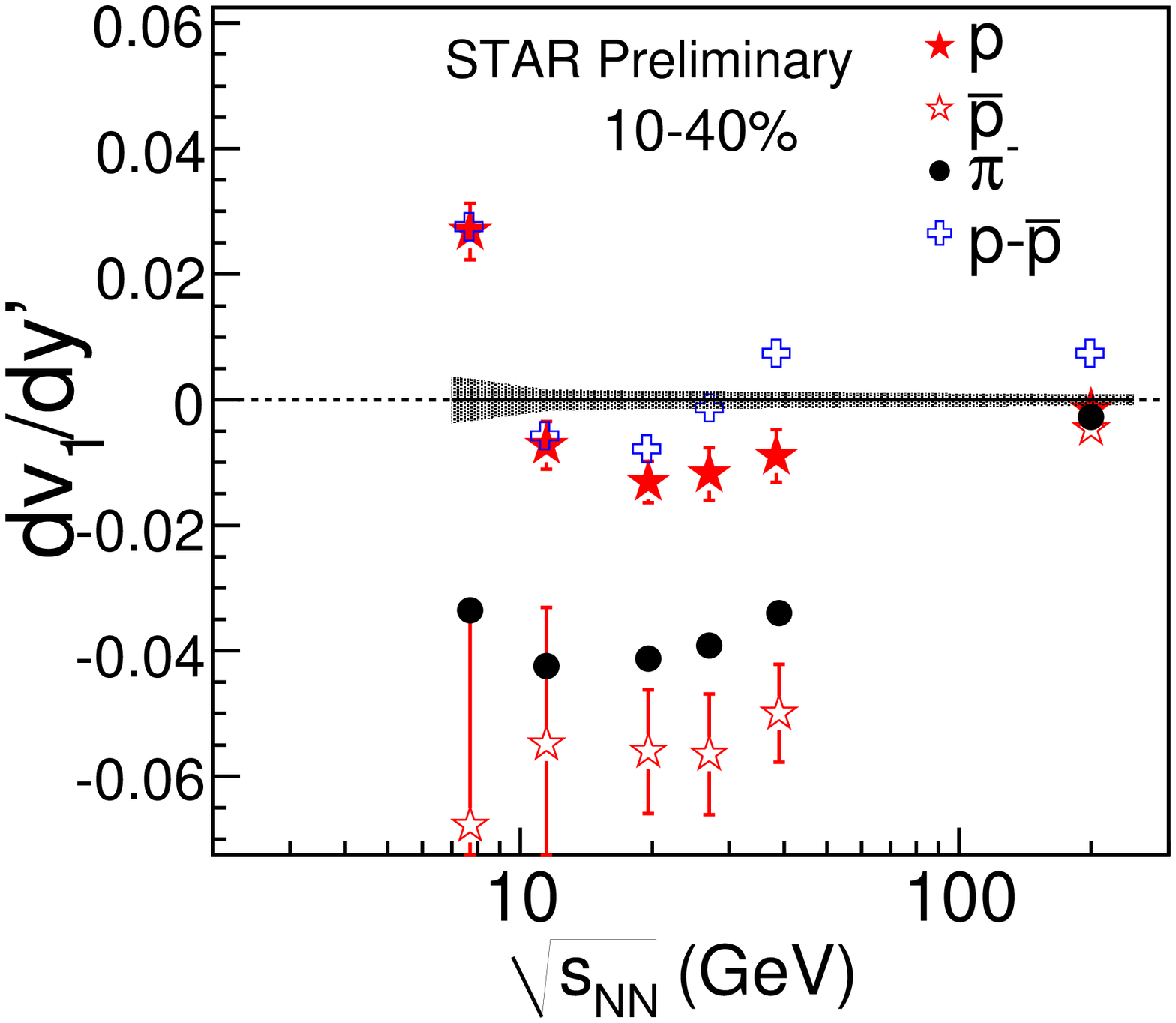}
\vspace{-0.5cm}
\caption{(Color online)
Directed flow slope
($dv_1/dy'$, $y'$=$y/y_{\rm{beam}}$) 
for $\pi^{-}$, $p$, $\bar{p}$, and net-protons ($p$-$\bar{p}$) near midrapidity as
a function of beam energy for mid-central (10--40\%) Au+Au
collisions. 
The shaded band refers to the systematic uncertainty on
net-proton measurements.
}
\label{fig_v1}
\end{minipage}
\hspace{0.4cm}
\begin{minipage}{0.48\linewidth}
\centering
\vspace{-0.38cm}
\includegraphics[width=1.0\textwidth]{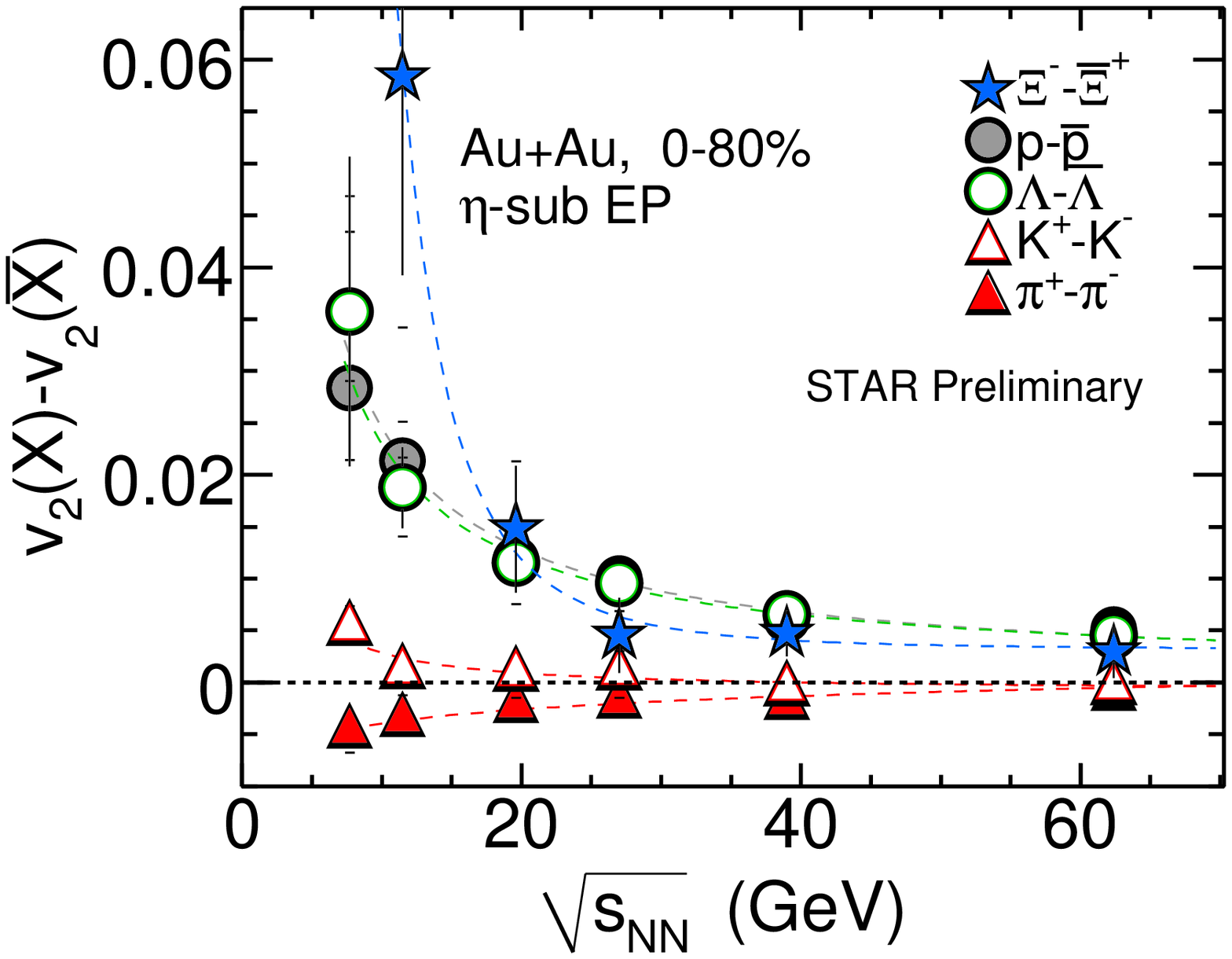}
\vspace{-0.65cm}
\caption{
(Color online) 
The difference in $v_2$ between particles
and their corresponding anti-particles as a function of beam energy in
0--80\% Au+Au collisions.
The curves represent fits to data points as discussed in text.
Both statistical (vertical lines) and
systematic errors (caps) are shown. 
}
\label{fig_v2}
\end{minipage}
\end{figure}
Figure~\ref{fig_v1} shows the $v_1$ slope ($dv_1/dy'$, where
$y'=y/y_{\rm{beam}}$ and $y$ is rapidity), near midrapidity as a function of beam energy for the mid-central
(10--40\%) Au+Au collisions~\cite{yadav}. The pion and anti-proton $v_1$ slopes show
negative values for all the beam energies studied. The proton $v_1$ slope
changes sign while going from 7.7 GeV to 11.5 GeV and then stays
negative up to 200 GeV. However, the net-protons 
$v_1$ slope 
(obtained using $v_1$ slopes of $p$, $\bar{p}$ and ratio of $\bar{p}/p$)
changes sign two times going from lower to
higher energy while showing a dip around $\sqrt{s_{NN}}$= 10--20 GeV. 
More studies are needed in order to understand these interesting observations.

\subsection{Elliptic Flow}
The elliptic flow $v_2$ is calculated as $\langle \cos2(\phi-\Psi_2)
\rangle$, where $\Psi_2$ is orientation of the second-order event plane.
Elliptic flow mainly probes the early stages of heavy-ion
collisions. 
At top RHIC energy of 200 GeV in Au+Au collisions, the elliptic flow scaled by the number
of constituent quarks ($n_q$) 
vs. ($m_T-m_0)/n_q$ (where $m_T=\sqrt{p_T^2 + m_0^2}$) 
shows a scaling behavior where mesons and baryons have similar values
at intermediate $p_T$.
This is referred to as the number of constituent quark (NCQ) scaling~\cite{ncq_ref}. It is an established
signature of partonic matter formed in Au+Au collisions at 200 GeV and deviations from
such scaling would indicate the
dominance of hadronic interactions.
Hence
breaking of NCQ scaling at lower energies could be an indication of a ``turn-off''
of QGP signatures. 
\begin{figure}[htbp]
\begin{center}
\vspace{-0.4cm}
\includegraphics[width=0.7\textwidth]{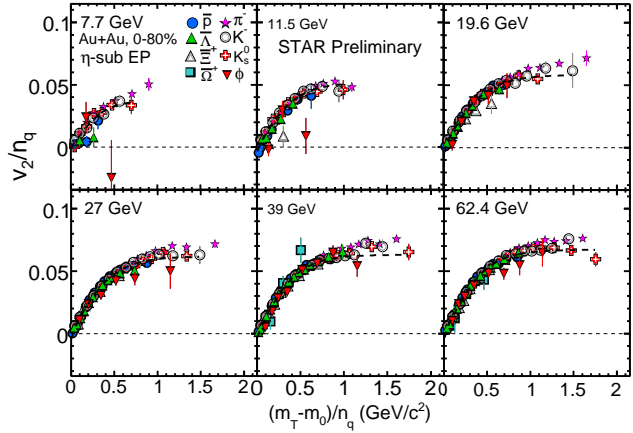}
\end{center}
\vspace{-0.75cm}
\caption{(Color online)  $v_2/n_q$ as a function
  of $(m_T-m_0)/n_q$ for different particles in Au+Au collisions at $\sqrt{s_{NN}}$= 7.7, 11.5, 19.6, 27, 39 and 62.4 GeV.
  The errors shown are statistical only.
}
\label{fig_ncq_v2}
\end{figure}
Figure~\ref{fig_v2} shows the
difference in $v_2$ of particles and corresponding anti-particles as a
function of beam energy~\cite{shusu}. 
The curves represent fits to data points with functional form:
$f_{\Delta v_2}(\sqrt{s_{NN}})=a \sqrt{s_{NN}}^{-b}+c $.
The $v_2$ difference between particles and
anti-particles is observed to increase when we go towards the lower
energies. At low energies, $v_2(\pi^{-}) > v_2(\pi^{+}$), $v_2(K^{+}) >
v_2(K^{-})$, and $v_2(\rm{baryons}) > v_2$(anti-baryons). This
difference between particles and anti-particles suggests that the NCQ
scaling among particles and anti-particles is broken. However, the
observed difference between $v_2$ of particles and anti-particles could
be qualitatively explained by the models incorporating baryon
transport at midrapidity and hadronic
interactions~\cite{ref_v2diff}. 
We also observe that the baryons-mesons splitting for $v_2$
vs. $m_T-m_0$  starts to disappear for
anti-particles at 11.5 GeV and below.
Figure~\ref{fig_ncq_v2} shows the $v_2/n_q$ vs. $(m_T-m_0)/n_q$ for different particles
for $\sqrt{s_{NN}}$= 7.7--62.4 GeV~\cite{shusu}. We observe that results for all the particles
are consistent among each other within $\pm10$\% level, except for the $\phi$-mesons. At the largest $m_T-m_{0}$ the $\phi$-meson data points deviate
by 1.8$\sigma$ and 2.3$\sigma$ for $\sqrt{s_{NN}}=$ 7.7 and 11.5 GeV,
respectively. 
Since $\phi$-mesons have smaller hadronic interaction
cross-section, their smaller $v_2$ values could indicate that the
hadronic interactions start to dominate over partonic effects for the
systems formed at beam energies below $\sqrt{s_{NN}}=$ 11.5
GeV~\cite{phi_bed}. However, as can been seen from the figure, a
higher statistics data are needed to extend the $m_T-m_{0}$ range
and significance of the deviation observed. 

\subsection{Dynamical Charge Correlations}
The dynamical charge correlations are studied through a three-particle
mixed harmonics azimuthal correlator~\cite{3partcorr},
$\gamma=\langle
cos(\phi_\alpha+\phi_\beta-2\Psi_{\rm{RP}})\rangle$. This observable
represents the difference between azimuthal correlations projected 
onto the direction of the angular momentum vector and correlations 
projected onto the collision reaction plane.
\begin{figure}[htbp]
\begin{center}
\includegraphics[width=0.8\textwidth]{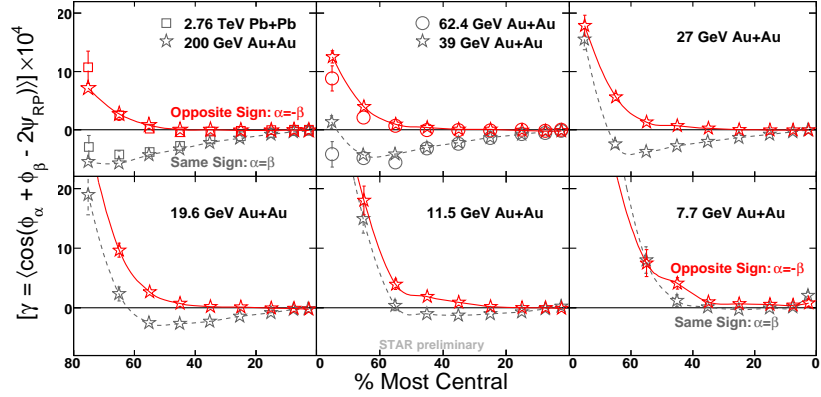}
\end{center}
\vspace{-0.5cm}
\caption{(Color online)
  $\langle \cos(\phi_\alpha+\phi_\beta-2\Psi_{\rm{RP}})\rangle$
as a function of centrality for Au+Au collisions at
$\sqrt{s_{NN}}=$7.7-200 GeV. For comparison, results for Pb+Pb
collisions at 2.76 TeV are also shown. Errors are statistical only.
}
\label{fig_dyn_ch}
\end{figure}
It is suggested that the difference in the correlations between same sign and
opposite sign charges in heavy-ion collisions could be related to local {\it parity}
violation if there is a deconfinement and a chiral phase
transition~\cite{lpv}. This is also referred to as Chiral Magnetic Effect
(CME). At top RHIC energies, we observed a separation between the
correlations of same and opposite sign charges.
If this difference can be
attributed to the QCD phase transitions, the absence of such observation
could be an indication of the system which did not undergo the phase
transition. Hence, the observable could be useful to locate the energy in
the BES program where the QGP signature ``turns off''.
Figure~\ref{fig_dyn_ch} shows the results for the beam energies from
7.7--200 GeV as a function of centrality~\cite{gang}. For comparison, Pb-Pb
results from ALICE are also shown~\cite{alice_lpv} 
which are
observed to be consistent with the results from top RHIC energy. The
separation between same and opposite sign charges decreases with 
decreasing energy and vanishes below $\sqrt{s_{NN}}=$11.5 GeV.

\subsection{Nuclear Modification Factor}
Another established signature of QGP at top RHIC energy is the Nuclear
Modification Factor $R_{\rm{CP}}$, which is defined as ratio of yields
at central collisions to peripheral collisions, scaled by the
corresponding number of
binary collisions. It has been observed that at high $p_T$, the
$R_{\rm{CP}}$ of different particles is less than unity~\cite{rcp}, which is
attributed to the energy loss of the partons in the dense medium. In
the absence of dense medium, there may not be suppression of high $p_T$
particles, which can serve as an indication of ``turn-off'' of a QGP
signature. 
\begin{figure}[htbp]
\begin{center}
\vspace{-1.0cm}
\includegraphics[width=0.75\textwidth]{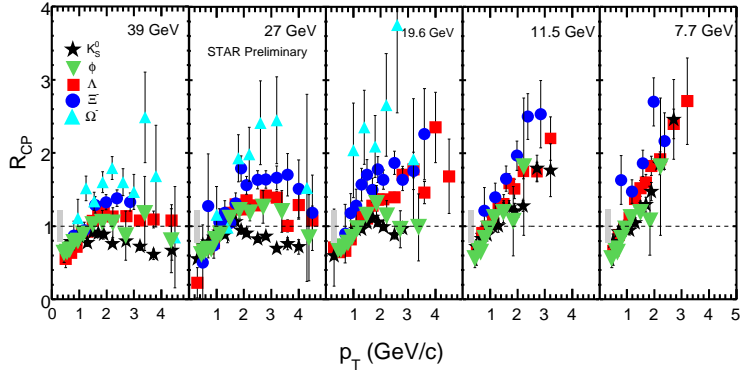}
\end{center}
\vspace{-0.75cm}
\caption{(Color online) $R_{\rm{CP}}$ (0--5\% / 40\%--60\%) for $K^{0}_{S}$, $\Lambda$, $\Xi^-$, 
  $\Omega^-$ and $R_{\rm{CP}}$ (0--10\% / 40\%--60\%)  for $\phi$-meson in Au+Au collisions at
  $\sqrt{s_{NN}}=$7.7--39 GeV.
Errors are statistical only.
Grey bands represent the normalization error from $N_{\rm{bin}}$.
}
\label{fig_rcp}
\end{figure}
Figure~\ref{fig_rcp} shows the $R_{\rm{CP}}$ of various strange
hadrons such as $K^{0}_{S}$, $\phi$, $\Lambda$, $\Xi^-$, and $\Omega^-$
in Au+Au collisions at $\sqrt{s_{NN}}=$7.7--39 GeV~\cite{xiaoping}. We observe
that for $p_T>$ 2 GeV/$c$, the $R_{\rm{CP}} (K^0_S)$  is less than
unity at 39 GeV and then the value increases as the beam energy decreases. For
$\sqrt{s_{NN}}<$ 19.6 GeV, $R_{\rm{CP}} (K^0_S)$ is above unity,
indicating decreasing partonic effects at lower energies. 
For (un)-identified charged hadrons results, please refer to~\cite{evan}.

\section{Search for QCD Critical Point}
If a system passes close to a critical point, the correlation length 
$\xi$ is expected to diverge. The higher moments (skewness $S$ and
kurtosis $\kappa$) of distributions of
conserved quantities such as net-baryons, net-charge, and
net-strangeness, have a better sensitivity to the correlation length
compared to variance $\sigma^2$~\cite{corr_length}. Also, the moment products such as
$\kappa\sigma^2$ and $S$$\sigma$ can be related to the ratios of order
susceptibilities calculated in Lattice QCD and HRG
model as 
$\kappa\sigma^2=\chi^{(4)}_B/\chi^{(2)}_B$ and
$S\sigma=\chi^{(3)}_B/\chi^{(2)}_B$~\cite{cp_lattice}. 
One of the advantages of using these
products is that they cancel the volume effects. Figure~\ref{fig_cp}
shows the $\kappa\sigma^2$ and $S\sigma$ for net-protons as a function
of beam energy for different collision centralities~\cite{xiaofeng}. 
For comparison, the results are shown for Poisson expectations and UrQMD model
calculations~\cite{urqmd}. 
The bottom panel
shows the $S\sigma$ values normalized by the corresponding Poisson
expectations. We observe that the moment products show similar values
for central 0--5\% and peripheral collisions for
$\sqrt{s_{NN}}=$39--200 GeV. For beam energies below 39 GeV, 
they have different values for central and peripheral
collisions. Their values are below 
Poisson expectations for $\sqrt{s_{NN}}>$ 7.7 GeV for 0--5\% central
collisions. However, for peripheral collisions, moment products values
are greater than Poisson expectations below 19.6 GeV. 
The UrQMD model calculations show a smooth monotonic behavior as a
function of collision energy. It may be noted that we have large
uncertainties for data points below 19.6 GeV that calls for higher
statistics data for these energies.
For net-charge and particle ratio fluctuation results, please refer to~\cite{daniel}.

\section{The BES Phase-II and STAR in Fixed Target Mode}

The BES Phase-I program from STAR has several interesting results
as discussed above. We have collected sizable data for Au+Au
collisions over a wide range of collisions energies from 7.7 to 39 GeV. 
However, to consolidate the findings from Phase-I, we need higher statistics
at lower energies, especially at 7.7 and 11.5 GeV. As we have discussed
above, statistics for several important observables, such as $\phi$-meson $v_2$ and
higher moments of net-protons distributions, are not sufficient to draw
quantitative conclusions. 
Also, 
there is more than a 100 MeV gap in $\mu_B$ between
11.5 and 19.6 GeV, so
having one more energy point around
15 GeV would allow to confirm the trends happening between 11.5 and
19.6 GeV.

\begin{minipage}{\textwidth}
 \begin{minipage}[b]{0.5\textwidth}
  \centering
  \includegraphics[width=1.0\textwidth]{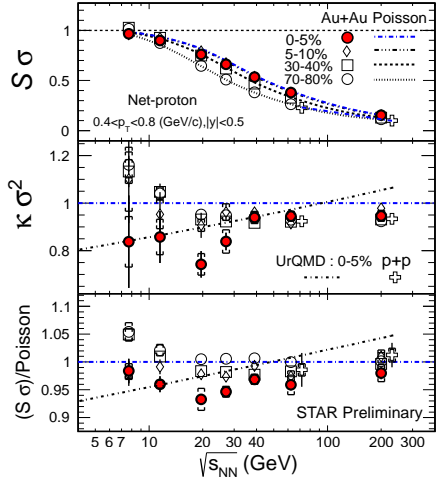}
\vspace{-0.8cm}
 \captionof{figure}{(Color online) $\kappa\sigma^2$, $S\sigma$ and $S\sigma$
 values normalized by the Poisson expectation, as a function of collision
 energy. Vertical bars are statistical and caps are systematic errors.
}
\label{fig_cp}
\end{minipage}
\hfill
\begin{minipage}[b]{0.5\textwidth}
 \centering
    \captionof{table}{The collider mode $\sqrt{s_{NN}}$, fixed target mode $\sqrt{s_{NN}}$,
and  corresponding $\mu_{B}$ values. The $\mu_B$ values are listed for
the central collisions corresponding to fixed target $\sqrt{s_{NN}}$~\cite{cleymans}.}
\begin{tabular}{c|c|c}
\hline
Collider mode & Fixed-target mode  & $\mu_{B}$   \\
 $\sqrt{s_{NN}}$ (GeV) &  $\sqrt{s_{NN}}$ (GeV) & (MeV) \\
\hline
19.6   & 4.5   & 585       \\
15  & 4.0   &  625 \\ 
11.5  &3.5   & 670     \\
7.7  &  3.0  & 720      \\
5  & 2.5   & 775      \\
\hline
\end{tabular}
\vspace{3cm}
   \label{table2}
 \end{minipage}
\end{minipage}
\vspace{0.5cm}

In view of these, STAR has
proposed a BES Phase-II program. For this STAR has requested an
electron cooling device at RHIC to increase the luminosity
for
collisions below $\sqrt{s_{NN}}=$ 20 GeV. Simulation results indicate
that with electron cooling, the luminosity could be increased by a
factor of about 3--5 at 7.7 GeV and about 10 around 20 GeV~\cite{ecool}. An
additional improvement in luminosity may be possible by operating with
longer bunches at the space-charge limit in the
collider~\cite{longbunch}. So overall improvement in luminosity could
be about 10-fold. The luminosity improvement will not only allow the
precision measurements of the above important observables but will
also be helpful in the measurements of rare probes such as dilepton
production and hypertriton measurements~\cite{rare}. 
To maximize the use of collisions provided at STAR for the BES program, we have proposed
to run STAR in ``fixed-target mode''. The main motivation for this proposal is
to extend the $\mu_B$ coverage currently from 400 MeV to about 800 MeV
so as to cover a large portion of the phase diagram. For this, we plan
to install a fixed Au target in the beam pipe to perform the Au(beam)-Au(target)
collisions. 
The data taking can be done concurrently during the normal RHIC
running. This proposal will not affect the normal RHIC
operations.
Table~\ref{table2} lists the proposed collision energies,
corresponding fixed target center-of-mass energies, and baryon chemical
potential values.

\section{Summary}
We have presented several interesting results from the STAR BES Phase-I
program. Currently, STAR covers a large range of $\mu_B$ (20--400 MeV)
in the phase
diagram. 
We have observed a centrality dependence of freeze-out
parameters at lower energies. We have looked at various observables for
the search of first order phase transition, ``turn-off'' of the QGP
signatures, and search of a QCD critical point.
The observables such as
$v_1$ slope, $R_{\rm{CP}}$ and $v_2$ of PID
hadrons, and higher moments of net-proton distributions, show some
dramatic changes for the 
energy range $\sqrt{s_{NN}}<$ 20 GeV.
We have observed that proton $v_1$ slope changes sign between
7.7 and 11.5 GeV. The net-proton $v_1$ slope changes sign twice
as a function of beam energy. 
We have observed a difference in $v_2$ of particles and corresponding
anti-particles which increases with decreasing collision energy. The
$\phi$-meson $v_2$ deviates from that of other particles,
the charge separation signal vanishes,
and the $R_{\rm{CP}}$ of $K^0_S$ is greater than unity for beam
energy of 11.5 GeV and below. 
The moment products have values below the Poisson
expectation for 0--5\% central Au+Au collisions. 
For peripheral collisions, they are above unity at $\sqrt{s_{NN}} <$
19.6 GeV. 
An electron
cooling device at RHIC will significantly increase luminosity at beam energies below
20 GeV
allowing the precision measurements for several important observables 
for phase structure studies.
In addition, we
propose to run STAR in fixed target mode which will extend
the $\mu_B$ coverage from STAR up to 800 MeV in the phase
diagram. 

We acknowledge the support from DOE for this research.


\end{document}